\documentclass[aps,floatfix,twocolumn]{revtex4}
\usepackage{graphics,epsfig}
\usepackage{graphicx}
\usepackage{amsmath,amssymb}
\usepackage{dcolumn}

\begin{document}

\title{Particle Accelerators inside Spinning Black Holes}
\author{Kayll Lake}
\email{lake@astro.queensu.ca}
\affiliation{Department of Physics, Queen's University, Kingston,
Ontario, Canada, K7L 3N6 }

\date{\today}

\begin{abstract}
On the basis of the Kerr metric as a model for a spinning black hole accreting test particles from rest at infinity, I show that the center-of-mass energy for a pair of colliding particles is generically divergent at the inner horizon. This shows that not only are classical black holes internally unstable, but also that Planck-scale physics is a characteristic feature within black holes at scales much larger than the Planck length. The novel feature of the divergence discussed here is that the phenomenon is present only for black holes with rotation and in this sense it is distinct from the well known Cauchy horizon instability.
\end{abstract}
\maketitle

\textit{Introduction.} - Recently, Ba\~{n}ados, Silk and West ~\cite{Banados:2009pr} (BSW) suggested that rotating black holes could serve as particle
colliders with arbitrarily high center-of-mass energies, possibly offering a visible probe of Plack-scale physics. This suggestion was soon criticized. Berti et al~\cite{Berti:2009bk} pointed out that the BSW mechanism requires very fine tuning (a degenerate (maximally spinning) black hole and a critical angular momentum for one of the accreted particles). Further, they pointed out that in the real world one would obtain only modest center-of-mass energies due to the Thorne upper limit on the angular momentum of a black hole \cite{Thorne:1974ve}. Moreover, they pointed out that the effects of gravitational radiation are not ignorable. At about the same time, Jacobson and Sotiriou~\cite{Jacobson:2009t} carefully analyzed the fine tuning required by the BSW mechanism, also pointed out the consequences of the Thorne limit, and showed how the redshift further lowers realizable energies. It would seem that the result obtained by BSW cannot be realized in nature, but the fact that arbitrarily high center-of-mass energies can in principle arise remains fascinating. However, for a maximally spinning black hole the inner (Cauchy) horizon coincides with the event horizon. Now since black hole Cauchy horizons have been known, for many years, to be unstable~\cite{Ori}, one might well suspect that this finely tuned divergence is somehow related to Cauchy horizon instability.

In this Letter I show that spinning black holes, approximated by the Kerr metric, do catalyze hyper-relativistic particle collisions, not about their outer horizons, but rather in the vicinity of their inner horizons. Moreover, I show that this divergence is a generic feature of rotating black holes in that the result requires no fine tuning at all. This instability is reminiscent of Cauchy horizon instability. However, it is distinct. Whereas Cauchy horizon instability remains in the limit of no rotation, the instability discussed here does not. To catalyze hyper-relativistic particle collisions inside black holes, we show that the black holes must have angular momentum. To emphasize the importance of the angular momentum, we motivate the  four-dimensional calculation by first considering a three-dimensional rotating black hole. We show that the instability is already present in three dimensions.

\textit{Methodology.} - For a pair of colliding particles (say $1$ and $2$) of (for simplicity) equal mass $m$, the center-of-mass energy is given by the covariant relation \cite{notation}
\begin{equation}\label{cm}
\left(\frac{E_{\rm cm}}{m}\right)^2=2(1-g_{\alpha \beta}u^{\alpha}_{(1)}u^{\beta}_{(2)})
\end{equation}
where $u^{\alpha}$ is the unit 4-velocity of the particle. In what follows we simply apply (\ref{cm}) in various backgrounds described by the metric $g_{\alpha \beta}$ with appropriate choices for the velocities $u^{\alpha}$. It is certainly a surprise that this straightforward calculation appears not to have been carried out in the Kerr metric prior to the BSW analysis.

\textit{No Rotation.} - In the absence of rotation we consider the static fields
\begin{equation}\label{static}
ds^2=-fdt^2+\frac{dr^2}{f}+r^2d\Omega^2_{2},
\end{equation}
where $f=f(r)$ and $d\Omega^2_{2}$ is the metric of a unit two-sphere ($d\theta^2+\sin^2 \theta d \phi^2$). These fields include the Reissner - Nordstr\"{o}m - de Sitter solutions to the Einstein equations. We consider the non-degenerate cases so the horizons occur at simple roots $f=0$ (say $r=r_{_{0}}$). The $t$ - independence of (\ref{static}) gives rise to a conserved energy $\gamma\; (\neq 0)$ for test particles and the $\phi$ - independence of (\ref{static}) gives rise to a conserved angular momentum $l$. Without loss in generality we set $\theta=\pi/2$. The resultant 4-velocities are given by
\begin{equation}\label{4v}
{u}^{a}=\left({\frac {\gamma}{f}},-\sqrt {{\gamma}^{2}-f\left( 1+{\frac {{l}^{2}}{{r}^{2}}} \right) },0,{
\frac {l}{{r}^{2}}}\right).
\end{equation}
From (\ref{cm}), (\ref{static}) and (\ref{4v}), after use of l'H\^{o}pital's rule, we find \cite{bswcase} \cite{labels}
\begin{equation}\label{statice}
 \left(\frac{E_{\rm cm}}{m}\right)^2_{r_{_{0}}}= {\frac {{r_{_{0}}}^{2} \left( \gamma_{{1}}+\gamma_{{2}} \right) ^{2}+
 \left( l_{{1}}\gamma_{{2}}-\gamma_{{1}}l_{{2}} \right) ^{2}}{\gamma_{
{1}}\gamma_{{2}}{r_{_{0}}}^{2}}}.
\end{equation}
We conclude that without rotation, the center-of-mass energy remains finite at non-degenerate horizons. This regularity contrasts with Cauchy horizon instability \cite{Ori}.

\textit{A Little Rotation.} - By a little rotation we mean a 2+1 dimensional rotating black hole, and not a black hole with a little rotation. The black hole considered here is the well known BTZ black hole \cite{BTZ}. The purpose here is to motivate a full four-dimensional calculation. The metric can be written in the form
\begin{equation}\label{BTZ}
ds^2=(-f+\frac{J^2}{4r^2})dt^2+\frac{dr^2}{f}-Jdt d\phi+r^2 d \phi^2,
\end{equation}
where $J$ represents the angular momentum and $f=f(r)$, the specific form of which does not concern us here. Again we consider the non-degenerate cases so the horizons occur at simple roots $f=0$ (say $r=r_{_{0}}$). The geodesic structure of the BTZ black hole is known \cite{cruz} and all we need to note here is that the turning points for timelike geodesics occur above the outer horizon and below the inner horizon. With the same notation as above, and considering the inner horizon, we now find a \textit{divergence} of the form
\begin{equation}\label{BTZE}
\left(\frac{E^{^{BTZ}}_{\rm cm}}{m}\right)^2_{r_{_{0}}} \sim \frac{(J l_1+2 \gamma_1 r_{_{0}}^2)(J l_2+2 \gamma_2 r_{_{0}}^2)}{-2 r_{_{0}}^4f(r_{_{0}})}.
\end{equation}
(Note that there is no double-horizon structure in the limit $J \rightarrow 0$ and so there is no $J \rightarrow 0$ limit for (\ref{BTZE}).)
We now explore this divergence more thoroughly in the four dimensional case.

\begin{figure}[ht]
\epsfig{file=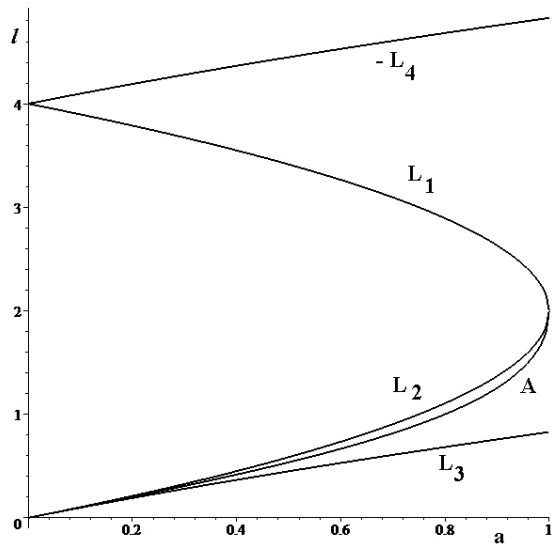,height=3in,width=3in,angle=0}
\caption{\label{L}The roots $L$ along with $A$ for comparison. The purpose of this diagram is to show that our choice for the range $L_2 <l_2 <L_1$ excludes the fine-tuning $l_2=A$ characteristic of the BSW mechanism.}
\end{figure}

\begin{figure}[ht]
\epsfig{file=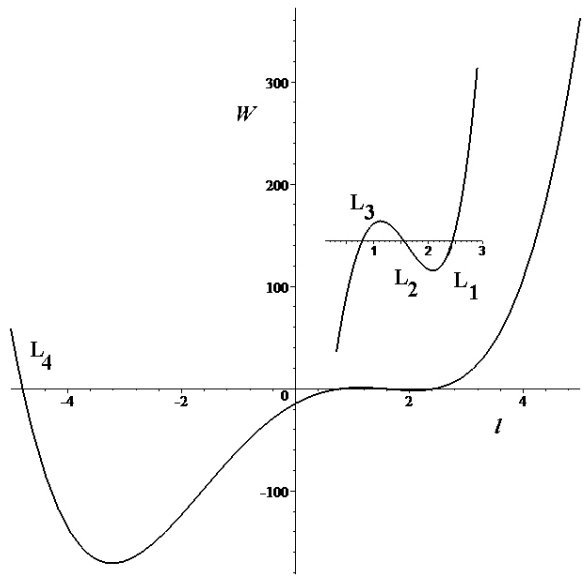,height=3in,width=3in,angle=0}
\caption{\label{ww}A sketch of the function $W$ (shown for $a=0.95$) with details in the insert. The roots $L$ are shown. The purpose of this diagram is to show that our choice for the range $L_4 <l_1 <L_3$ does not require counter-streaming, that is, $l_1 <0$.}
\end{figure}

\begin{figure}[ht]
\epsfig{file=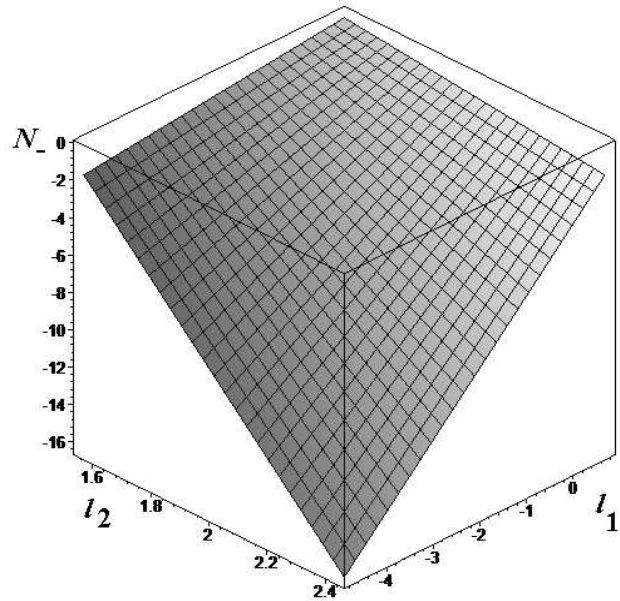,height=3.5in,width=3.5in,angle=0}
\caption{\label{Nfig}A plot of $N_{-}$, given by (\ref{EcmKs}), for $a=0.95$ within the stated ranges for $l_{1}$ and $l_{2}$. Other plots in the range $0<a<1$ are qualitatively similar. The purpose of this diagram is to show that $N_{-}$ does not evaluate to $0$ generically.}
\end{figure}

\textit{Rotation.} - As shown in~\cite{Banados:2009pr} and~\cite{Jacobson:2009t}, for a pair of particles of mass $m$ that fall from rest at infinity ($\gamma=1$) in the equatorial plane, the center-of-mass energy in the Kerr metric is given by
\begin{equation}\label{E}
\left(\frac{E^{^{Kerr}}_{\rm cm}}{m}\right)^2=\frac{2 N}{r (r^2 - 2r + a^2)}
\end{equation}
where
\begin{eqnarray}
N=2 a^2 (1 + r)-2 a (l_1 + l_2) - l_1 l_2 (r-2) + 2 (r-1) r^2\;\;\nonumber\\
\nonumber\\
-\sqrt{(2 (a - l_1)^2 - l_1^2 r + 2 r^2)(2 (a - l_2)^2 - l_2^2 r + 2 r^2)}\;,\;\label{EcmK}
\end{eqnarray}
the black hole is given unit mass, the angular momentum per unit mass of the black hole is given by $a$ and the particles have orbital angular momenta of $l_{1}$ and $l_{2}$ \cite{limit}. Here we consider black holes in the range $0<a<1$.
The horizons are given by $r_{\pm}\equiv1\pm\sqrt{1-a^2}$ and here we are concerned only with the inner horizons $r_{_{-}}$. To prove the general divergence at $r_{_{-}}$ first note that the denominator of $E^{^{Kerr}}_{\rm cm}$ obviously vanishes there. For the numerator we note that $N$ evaluates to
\begin{eqnarray}
N_{-}=-2a(l_{{1}}+l_{{2}})+l_{{1}}l_{{2}}r_{_{+}}+4r_{_{-}}\;\;\;\;\nonumber\\
\nonumber\\
-\sqrt{({l_{{1}}}^{2}r_{_{+}}+4(r_{_{-}}-\,al_{{1}})) ({l_{{2}}}^{2}r_{_{+}}+4(r_{_{-}}-\,al_{{2}}))}\;\;\label{EcmKs}
\end{eqnarray}
at $r=r_{_{-}}$.
 Whereas the detailed properties of geodesics in the Kerr metric are involved \cite{Kraniotis}, it is adequate for our purposes here to note the following: In the Kerr metric, timelike geodesics that fall from rest at infinity in the equatorial plane satisfy
\begin{equation}\label{rdot}
r^3 \dot{r}^2=2r^2-l^2r+2l^2-4la+2a^2
\end{equation}
and so $r^3 \dot{r}^2>0$ for
\begin{equation}\label{W}
    -(4a-4l+l^2)(4a-4l-l^2) \equiv W < 0.
\end{equation}
Now $W=0$ at 4 values of $l$, given by
\begin{equation}\label{L12}
L_{1,2}=2(1 \pm \sqrt{1-a}),
\end{equation}
and
\begin{equation}\label{L34}
L_{3,4}=-2(1 \mp \sqrt{1+a}).
\end{equation}
We note that $L_1  \rightarrow L_2 \rightarrow 2$ as $a \rightarrow 1$ and $L_2  \rightarrow L_3 \rightarrow 0$ as $a \rightarrow 0$. These roots are shown in Fig \ref{L} along with
\begin{equation}\label{A}
A={\frac {2a}{1+\sqrt { \left( 1-a \right)  \left(1+a \right) }}}
\end{equation}
for comparison. There is a critical turning point (inflexion in $r$) at $r=r_{_{-}}$ for $l=A$. The fine tuning in the BSW mechanism is due to the fact that $L_{1}=L_{2}=A$ at $r=r_{_{-}}$ for $a=1$ as discussed at length in \cite{Jacobson:2009t} and as shown in Fig \ref{L}.
A sketch of $W$ is shown in Fig \ref{ww}.
Over the range $0<a<1$ particles with
\begin{equation}\label{l1range}
   L_4<l_{1}<L_3
\end{equation}
and
\begin{equation}\label{l2range}
    L_2<l_{2}<L_1
\end{equation}
have no turning points.
Over the stated ranges in $l_{1}$ and $l_{2}$, $N_{-}$ does not evaluate to zero. This is demonstrated in Fig ~\ref{Nfig}.
It is important to note that the divergence discussed here is not due to any fine tuning (e.g. $l=A$ lies outside the range chosen for $l_{2}$), nor is it due to any requirement of counter-steaming (that $l_{1}$ and $l_{2}$ have opposite signs (they do not need to)).

\textit{Discussion.} - Because of the generic nature of the divergence discussed above, a divergence that suffers none of the limitations of the BSW mechanism, it is reasonable to conclude that the use of the Kerr metric and the test point-particle geodesic approximation has not given rise to a fictitious result. Given this, the principal conclusion here is that Planck-scale physics is a characteristic feature of black hole interiors at scales much larger than the Planck length. Further, we have found that the instability examined here is reminiscent of but quite distinct from the well-known Poisson-Israel instability \cite{Ori}. The instability of black hole Cauchy horizons does not require non-zero angular momentum of the black holes. The instability discussed here does. We note that the instability discussed here is already present in 2+1 dimensional rotating black holes. Finally, let us look at some of the simplifications used in the present argument for a four-dimensional spinning black hole. Whereas we have considered motion only in the equatorial plane, continuity strongly suggests that the divergence discussed here is not restricted to particle motion in the equatorial plane alone. We have used the test point-particle geodesic approximation. It would be interesting to see if the relaxation of this approximation changes the divergence.

\bigskip

\textit{Acknowledgments.} - This work was supported by a grant from the Natural Sciences and Engineering Research Council of Canada. Portions of this work were made possible by use of \textit{GRTensorII} \cite{grt}.


\begin{thebibliography}{}
\bibitem{notation}We use geometrical units and a signature of $+2$.
\bibitem{labels}
    The labels $1$ and $2$ are, of course, interchangeable.
\end{thebibliography}
\end{document}